\documentclass[10pt,journal]{IEEEtran}

\usepackage{graphicx}
\usepackage{subfig}
\usepackage{listings}
\usepackage{enumitem}

\graphicspath{{../images/}}
\DeclareGraphicsExtensions{.pdf,.jpeg,.png,.jpg,.eps}

\begin{document}
%
\title{The BigDAWG Architecture}



%
\author{\IEEEauthorblockN{Vijay
    Gadepally\IEEEauthorrefmark{1}\IEEEauthorrefmark{2} Jennie
    Duggan\IEEEauthorrefmark{3} Aaron Elmore\IEEEauthorrefmark{4},
    Jeremy Kepner\IEEEauthorrefmark{1}\IEEEauthorrefmark{2}
    Samuel Madden\IEEEauthorrefmark{2} Tim
    Mattson\IEEEauthorrefmark{5} Michael
    Stonebraker\IEEEauthorrefmark{2}} \\
\IEEEauthorblockA{\IEEEauthorrefmark{1}MIT Lincoln Laboratory \IEEEauthorrefmark{2}MIT CSAIL \IEEEauthorrefmark{3}Northwestern
  University \IEEEauthorrefmark{4}University of Chicago \IEEEauthorrefmark{5}Intel}}

\maketitle

\begin{abstract}

BigDAWG is a polystore system designed to work on complex problems
that naturally span across different processing or storage
engines. BigDAWG provides an architecture that supports diverse
database systems working with different data models, support for the
competing notions of location transparency and semantic completeness
via \textit{islands of information} and a middleware that
provides a uniform multi--island interface. In this article, we describe the
current architecture of BigDAWG, its application on the MIMIC II
medical dataset, and our plans for the mechanics of cross-system
queries. During the presentation, we will also deliver a brief
demonstration of the current version of BigDAWG.
\end{abstract}


\IEEEpeerreviewmaketitle

\section{Introduction}

Enterprises today encounter many types of databases, data, and
storage models. Developing analytics and applications that
work across these different modalities is often limited by the
incompatibility of systems or the difficulty of creating new
connectors and translators between each one. For example, consider the MIMIC II
dataset~\cite{saeed2011multiparameter} which contains deidentified
health data
collected from thousands of critical care patients in an Intensive
Care Unit (ICU). This publicly available dataset (\textit{http://mimic.physionet.org/})
contains structured data such as demographics and medications; unstructured text
 such as doctor and nurse reports; and time--series data
of physiological signals such as vital signs and electrocardiogram
(ECG). Each of these components of the dataset can be efficiently organized into
different styles of database engines. For example, the structured data
in a relational database, the text notes in a key-value or graph
database and the time--series data in an array database.  
Analytics of the future will cross the boundaries of a single data
modality, such as
correlating information from a doctor's note against the physiological
measurements collected from a particular sensor. Such analytics call for the development of a new generation of
federated databases that support access to different styles of
database or storage engines. We refer to such a system as a \textit{polystore}
 in order to distinguish it from traditional federated
databases that largely supported access to multiple
engines using the same data model.   



As a part of the Intel Science and Technology Center (ISTC) on Big
Data, we are developing BigDAWG, short for Big Data Analytics Working Group. The
BigDAWG stack is designed to support many sizes, real-time streaming
analytics, visualization interfaces, and multiple databases. The
current version of BigDAWG~\cite{elmore2015demonstration} shows
significant promise and has been used to develop a series of 
applications using the MIMIC II dataset. In this presentation, we will
describe the current BigDAWG architecture and
describe a series of polystore workloads developed for the MIMIC II dataset. We also describe our plans for the mechanics of
implementing cross-system queries. 

\section{BigDAWG Architecture}

\begin{figure}[b!]
\centerline{
\includegraphics[width=3in]{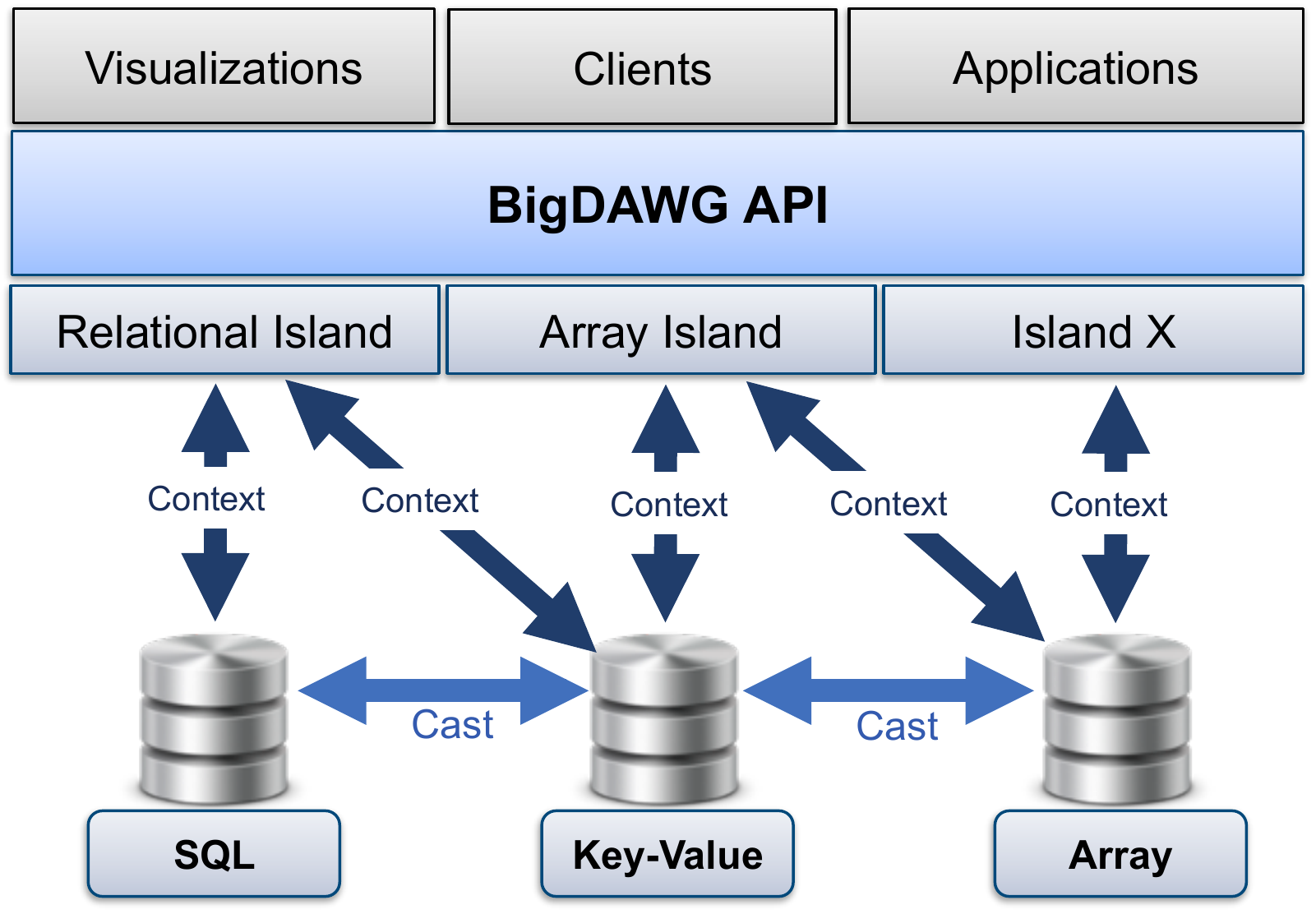}
}
\caption{The BigDAWG architecture.}
\label{fig:bigdawgarch}
\end{figure}

The BigDAWG architecture consists of four distinct layers as described
in Figure~\ref{fig:bigdawgarch}: database and storage engines; islands
of information; API; and applications. In this section, we discuss the
current status of each of these layers as well as how they are used
with the MIMIC II dataset.

\subsection{Database and Storage Engines}

A key design feature of BigDAWG is the support of multiple
database and storage engines. We believe that analytics of the future
will depend on many, disparate data sources and BigDAWG is designed to
address this need by leveraging many vertically-integrated data management systems.

For the MIMIC II dataset, we use the relational databases PostgreSQL and
Myria~\cite{halperin2014demonstration} to store clinical data such
as demographics and medications. BigDAWG uses the
key-value store Apache Accumulo for freeform text data and to perform graph analytics~\cite{hutchison2015graphulo}. For the historical
waveform time-series data of various physiological
signals, we use the array store
SciDB~\cite{stonebreaker2013}. Finally, for streaming timeseries data,
our application uses the streaming database S-Store~\cite{cetintemel2014s}.

\subsection{Islands of Information}

The next layer of the BigDAWG stack is its \textit{islands of
information}. Islands allow users to trade off between semantic
completeness (using the full power of an underlying database engine)
and location transparency (the ability to access data without
knowledge of the underlying engine). Each island has a data model, a
query language or set of operators and one or more database
engines for executing them. In the BigDAWG prototype, the user determines
the {\it scope } of their query by speficying an island of information
within which the query will be executed.
 Islands are a user-facing abstraction, and they are designed to reduce the challenges associated with incorporating a new database engine. 

We currently support a number of islands. For example, the D4M island
provides users with an associative array data model~\cite{vijay2015} to PostgreSQL,
Accumulo, and SciDB. The Myria island provides iteration to the
MyriaX, PostgreSQL and SciDB databases. We also support a number of 
\textit{degenerate} islands that connect to a single database engine. \textit{Degenerate} islands provide support for
the full semantic power of a connected database at the expense of
location transparency.

\vspace{-10pt}

\subsection{BigDAWG API}

The BigDAWG interface provides a simple API to execute polystore
queries. The API layer consists of server and client facing
components. The server components incorporate the many possible islands which connect
to database engines via lightweight connectors referred to as \textit{shims}. The clients interact with these islands
through the API via SCOPE and CAST operations. To specify a particular island of
information, a user indicates a SCOPE in their query. A SCOPE allows users to control
the data model and programming interface with which they wish their queries to
be executed. A cross-island query may be composed of multiple SCOPEs. For queries that rely on cross-island
interaction, BigDAWG also offers CAST operations that can
automatically move data
between database and storage engines -- and subsequently between
islands. We discuss the mechanics of these operations in greater
detail in Section~\ref{querymechanics}.

\vspace{-10pt}

\subsection{Applications and Visualizations}

BigDAWG supports a variety of visualization platforms such as
Vega~\cite{satyanarayan2016reactive} and
D3~\cite{bostock2011d3} which can be used to develop complex
applications and analytics. The current BigDAWG implementation was applied to the MIMIC II
medical dataset to develop different applications and analytics that
require polystore support. The applications developed were:

\begin{enumerate}[leftmargin=*]
\item \textbf{Browsing}: This screen provides an interface to the full
  MIMIC II dataset which is stored in different storage engines. This
  screen utilized the open source tool ScalaR~\cite{battle2013dynamic}.
\item \textbf{``Something interesting''}: This application uses
  SeeDB~\cite{vartak2014seedb} to highlight interesting trends and anomalies in the dataset. 
\item \textbf{``Text Analytics''}: This application performs topic modeling
  of the unstructured doctor and nurse notes directly in a key-value store
  database using Graphulo~\cite{hutchison2015graphulo}.
\item \textbf{``Heavy Analytics''}: This application looks for
  hemodynamically similar patients in a dataset by comparing the
  signatures of historical ECG waveforms using Myria.
\item \textbf{``Streaming Analytics''}: This application performs
  analytics on streaming time-series waveforms and can be used for ETL
  into another database such as SciDB. 
\end{enumerate}

\section{Executing Polystore Queries}
\label{querymechanics}

Efficient query execution is a key goal of the BigDAWG
system.  This aim is challenging because the data being queried is likely to be distributed among two or more disparate data management systems.  In order to support different islands, efficient data movement is essential.
Moreover, efficient execution depends on system parameters
such as available resources or usage that are prone to change. In this
section, we describe the simplest case where there is no replication, partitioned objects,
expensive queries or attempts to move objects for load balancing. An
execution plan for a query is then generated based on whether
the query is in a training or production phase.

\vspace{-10pt}

\subsection{Training Phase}

Training mode is typically used for execution of queries that are
new (either the query is new or the system has changed significantly
since the last time a particular query was run) or are believed to
have been poorly executed. In the simplest case, the training phase consists of queries that
arrive with a ``training'' tag. In the training phase, we allow the
query execution engine to generate a good query plan using any number
of available resources. First, the query preprocessor parses the query and scopes each piece of the query to a particular
island. Pieces of the resulting subquery that are local to a particular storage
engine are encapsulated into a container and given an identifying
signature. For the remaining elements of the query (remainder), which correspond
to cross-system predicates, we can generate a signature by looking at
the structure of the remainder, the objects being reference and the
constants in the query. If the remainder signature has been seen
before, a query plan can be extracted. If not, the system decomposes
the remainder to determine all possible query plans which are then
sent to the monitor.

To execute the query, the monitor feeds the queries to the executor,
plus all of the containers which are then passed to the appropriate
underlying storage engine. For the cross-engine predicates, the
executor decides how to perform each step. The executor runs each
query, collects the total running time and other usage statistics and
stores the information in the monitor database. This information can
then be used to determine the best query plan in the production phase.

\subsection{Production Phase}

In the production phase, when a query is received it is first matched
against the various signatures in the monitor database and the optimizer selects the closest
one. The BigDAWG optimizer also compares the current usage statistics of the system
and compares it against the usage statistics of the system when the
training was performed. If there are large differences, the optimizer
may select an alternate query plan that more closely resembles the
current resources or system usage or recommend that the user rerun the
query under 
the training phase under the current usage. In cases where the
signature of the incoming query do not match with existing
signatures, the optimizer may suggest the query run in training mode
or construct a list of plans as done in the training phase and have
the monitor pick one at random. The remaining plans can then be run in
the background of the system when it is underutilized. Over time,
these plans are added to the monitor database.


\section{Conclusion}
In this article, we described the latest architecture of the BigDAWG
system. A prototype version of the architecture has been built and
applied to the MIMIC II dataset. We described a few of these
applications and discuss our plan of attack for executing more complex
cross-system and cross-island queries.

\bibliographystyle{IEEEtran}
\bibliography{IEEEabrv,references.bib}

\end{document}